\documentclass[journal, final, 10pt, twocolumn]{IEEEtran}

\usepackage{graphicx}
\usepackage{amsmath}
\usepackage{mathtools}
\usepackage{epsfig}
\usepackage{float}
\usepackage{lipsum}
\usepackage{stfloats}
\usepackage{array}
\usepackage{amssymb}
\usepackage{cite}
\usepackage{url}
\usepackage{hyperref}
\usepackage{amsthm}
\usepackage{algorithm}
\usepackage{algorithmic}
\usepackage{multirow}
\usepackage{upgreek}
\usepackage{diagbox}
\usepackage{dsfont}
\usepackage{xifthen}% provides \isempty test
\usepackage{xcolor}
\usepackage[colorinlistoftodos,textsize=tiny]{todonotes}

\usepackage{caption}
\usepackage{subcaption}
\allowdisplaybreaks
\usepackage{layout}
\usepackage{bm}
\usepackage{comment}
\usepackage{color,soul}
\usepackage{makecell}
\usepackage{amssymb}

\begin{document}

%\title{Microservice Scaling Using MEC Sandbox and ETSI MEC APIs}
\title{Mobile Edge Vertical Applications Using ETSI MEC APIs and Sandbox}
%\author{\IEEEauthorblockN{Rasoul Nikbakht} \\
%	\IEEEauthorblockA{Centre Tecnològic Telecomunicacions Catalunya (CTTC) \\ 08860 Castelldefels, Spain\\
%		Email: rnikbakht@cttc.es}

%}

\author{\IEEEauthorblockN{Rasoul Nikbakht\IEEEauthorrefmark{1},
Michail Dalgitsis\IEEEauthorrefmark{2}, Sergio Barrachina-Mu\~noz\IEEEauthorrefmark{1}, Sarang Kahvazadeh\IEEEauthorrefmark{1}}

\IEEEauthorblockA{\IEEEauthorrefmark{1}Centre Tecnològic Telecomunicacions Catalunya (CTTC), Castelldefels, Spain}

\IEEEauthorblockA{\IEEEauthorrefmark{2}Vicomtech, San Sebastian, Spain} \\
rnikbakht@cttc.es, mdalgitsis@vicomtech.org, sbarrachina@cttc.es, skahvazadeh@cttc.es
}

\maketitle

\begin{abstract}
MEC Sandbox is an excellent tool that simulates wireless networks and deploys ETSI Multi-access Edge Computing (MEC) APIs on top of the simulated wireless network. In this demo, we consume these APIs using a decision engine (DE) to scale a video-on-demand (VoD) application located on the network edge, assuming that the average number of users is a good proxy of the demand. Specifically, the developed DE uses the ETSI MEC Location API and retrieves the number of users in a given zone. The DE then takes actions at the microservice scaling level and executes them through a custom-made Kubernetes-based OpenAPI. 

%\michail{cloud-native approach and API calls are quickly gaining ground among Telecom infrastructure providers and mobile operators. Leveraging the interactive environment of ETSI MEC Sandbox, we experiment with the ETSI MEC Location API and a custom-made Kubernetes-based OpenAPI to scale and monitor a video-on-demand (VoD) server, located on the network edge.}  

\end{abstract}

%------------------------------------------------------------
%------------------------------------------------------------
\section{Introduction}

% Introducing MEC APIs

Within the broad topic of edge computing, Multi-access Edge Computing (MEC) is the widely accepted standard for edge computing created by the European Telecommunications Standards Institute (ETSI). MEC provides computing and storage resources at the edge of 5G networks to let application and mobile operators reduce delay and backhaul traffic. To optimize the utilization of MEC resources, ETSI designed MEC APIs that provide real-time access to radio network information. The MEC APIs allow the operator to open their radio access network (RAN) to external users and share information such as location and trajectory with third-party applications, providing added value for operators and applications and improving service quality for end-users \cite{ETSI_mec_API}.

%\michail{Multi-access edge computing (MEC) provides computing and storage resources at the edge of 5G networks. Content providers and mobile operators can use MEC resources to reduce delay and backhaul traffic. To optimize the utilization of MEC resources, ETSI standardized OpenAPis called MEC Service APIs for the mobile and edge network. These APIs allow the operator to open their radio access network (RAN) and MEC sites to unauthorized users and share information such as location and trajectory with third-party applications, providing added values for operators and applications and improving service quality for end-users \cite{ETSI_mec_API}.}

% MEC Sandbox

To enable users to learn and experiment with MEC APIs, ETSI developed the MEC Sandbox, an interactive environment that provides a variety of scenarios mixing different network technologies and terminal types\cite{mec_sandbox}. The MEC Sandbox simulates a part of the city of Monaco with a street map, user mobility (pedestrians and vehicle users), and cell zones.
%\michail{I would put the paragraphs about MEC Sandbox in another section called discussions after the Experiment section}
%MEC sandbox is a tool developed by ETSI to facilitate application development with MEC APIs, producing a simulated environment \cite{mec_sandbox}. The MEC sandbox simulates a part of a city with a realistic base station deployment, street map, and user mobility (pedestrians and vehicle users). One can interact with users, RAN, and edge resources, using MEC APIs like location services and radio network information. % The MEC sandbox does not have any 5G core or RAN deployment. Rather, it provides a coherence API call that can be used to develop a realistic application.
%In this work, we use the MEC sandbox for location-based microservice scaling. We deploy a video-on-demand (VoD) streaming application at the edge and use the average number of users in the given zone of the MEC sandbox as a proxy for requested traffic for the VoD application.  
In this work, we leverage the MEC Sandbox assuming the simulated users stream a cloud-native video running at the MEC site, through a mobile network, by calling the MEC Location API to calculate the average number of users in a given zone. Should the traffic demand increase, the DE scales the video server and exports metrics to the monitoring system. 

%------------------------------------------------------------
%------------------------------------------------------------
\section{System architecture}
In this demonstration, we showcase the complete pipeline of an edge microservice scaling solution using MEC Sandbox and a custom-made Kubernetes OpenAPI. First, we deploy a Kubernetes cluster consisting of cloud and edge nodes. %Then, we create a cloud-native 5G network by deploying the Open5GS core in the cloud node and VoD streaming application alongside the user plane function (UPF) in the edge node.
Then, we use a custom Helm package~\cite{barrachina2022cloud,barrachina2022intent} to deploy a cloud-native 5G network realized with Open5GS network functions (NFs) that span cloud and edge domains. In particular, control plane NFs (e.g., AMF or SMF) run in the cloud, whereas the user plane function (UPF) runs in the edge node alongside a VoD streaming application. In addition, we deploy  \textit{i}) an ingress controller in the Kubernetes edge node for distributing the traffic between the pods of the VoD application \cite{Nikb2211:Video},  \textit{ii}) the custom-made OpenAPI server, \textit{iii}) the DE instance, and \textit{iv}) a  monitoring namespace, which includes Prometheus and Grafana pods to monitor and visualize the cluster, respectively.

%\michail{In this demonstration, we showcase edge microservice scaling and monitoring using the MEC sandbox and two OpenApis, the MEC Location API by ETSI, and a custom-made Kubernetes OpenAPI. The MEC site is deployed as a Kubernetes cluster consisting of two nodes. In the nodes are running the following pods: i) a cloud-native 5G core, using the Open5Gs solution, ii) the VoD streaming application, iii) the custom-made OpenAPI server, and iv) the controller. In addition, we have deployed in the monitoring namespace Prometheus and Grafana pods to monitor and visualize the cluster respectively.}

%MEC sandbox interface is shown in Fig. \ref{fig:mecsandboxfull}. We can specify the simulation network (4G-5G-wifi-macro) and the number of low-velocity or high-velocity users. Once the simulated network instance is up and running, we initialize the location service and deploy the decision engine in the cloud node according to Fig. \ref{fig:shematicoverall}. The decision engine interacts with the MEC sandbox using ETSI MEC APIs (in this case, location API), makes a threshold or machine learning (ML)-based decision, and finally, performs micro service scaling using Kubernetes python API. We also use the Prometheus Python client inside the decision engine to expose the relevant information to the monitoring system.

\begin{figure}
	\centering
	\includegraphics[width=.8\linewidth]{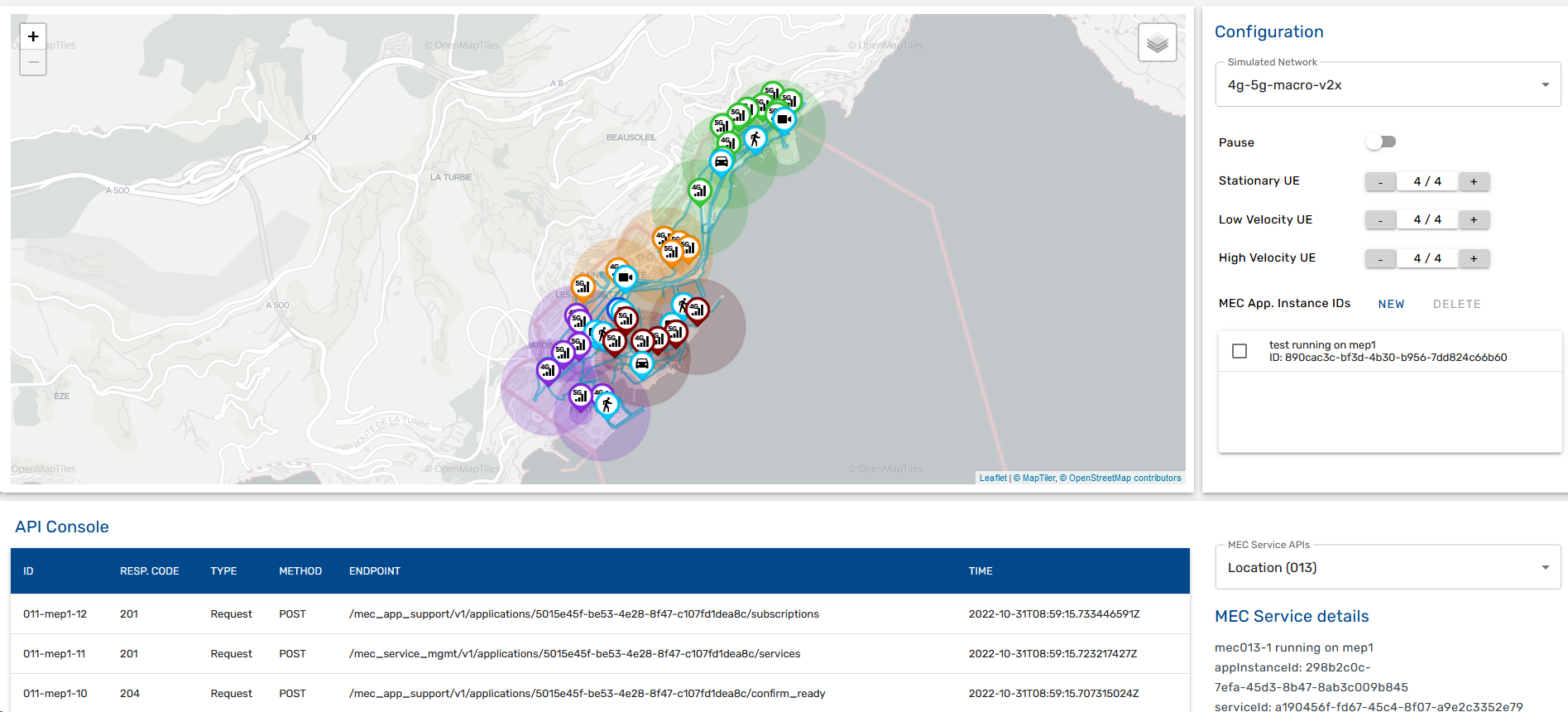}
	\caption{MEC Sandbox}
	\label{fig:mecsandboxfull}
\end{figure}

%Looking the MEC sandbox interface as it is shown in Fig. \ref{fig:mecsandboxfull},
As shown in the MEC Sandbox interface in Fig. \ref{fig:mecsandboxfull}, the network scenario and the number of users can be selected from the dashboard. Once the simulated network instance is up and running, the DE starts calling periodically the MEC Location API to find out the average number of users in a given zone. The DE, apart from integrating with MEC Sandbox, has various other roles such as exposing the collected metrics from the MEC Location API to the Prometheus server, acting as the custom-made OpenAPI client to request a scaling request, and defining the threshold for the triggered actions. To demonstrate the integration of APIs to cloud-native infrastructures, we build a custom-made swagger OpenAPI server with Kubernetes endpoints, providing an extra layer of abstraction to manage the cluster. This 2-tier API interaction gives an extra degree of freedom to automatically and remotely manage resources on the edge. 

%\michail{I would put this on the discussion section: Although Kubernetes natively supports horizontal pod autoscaling (HPA), our experiment goes beyond the current methods, allowing the opportunity for artificial intelligence algorithms to be applied in the controller and make predictions regarding network resource optimization}
 
{\begin{figure}
	\centering
	\includegraphics[width=.8\linewidth]{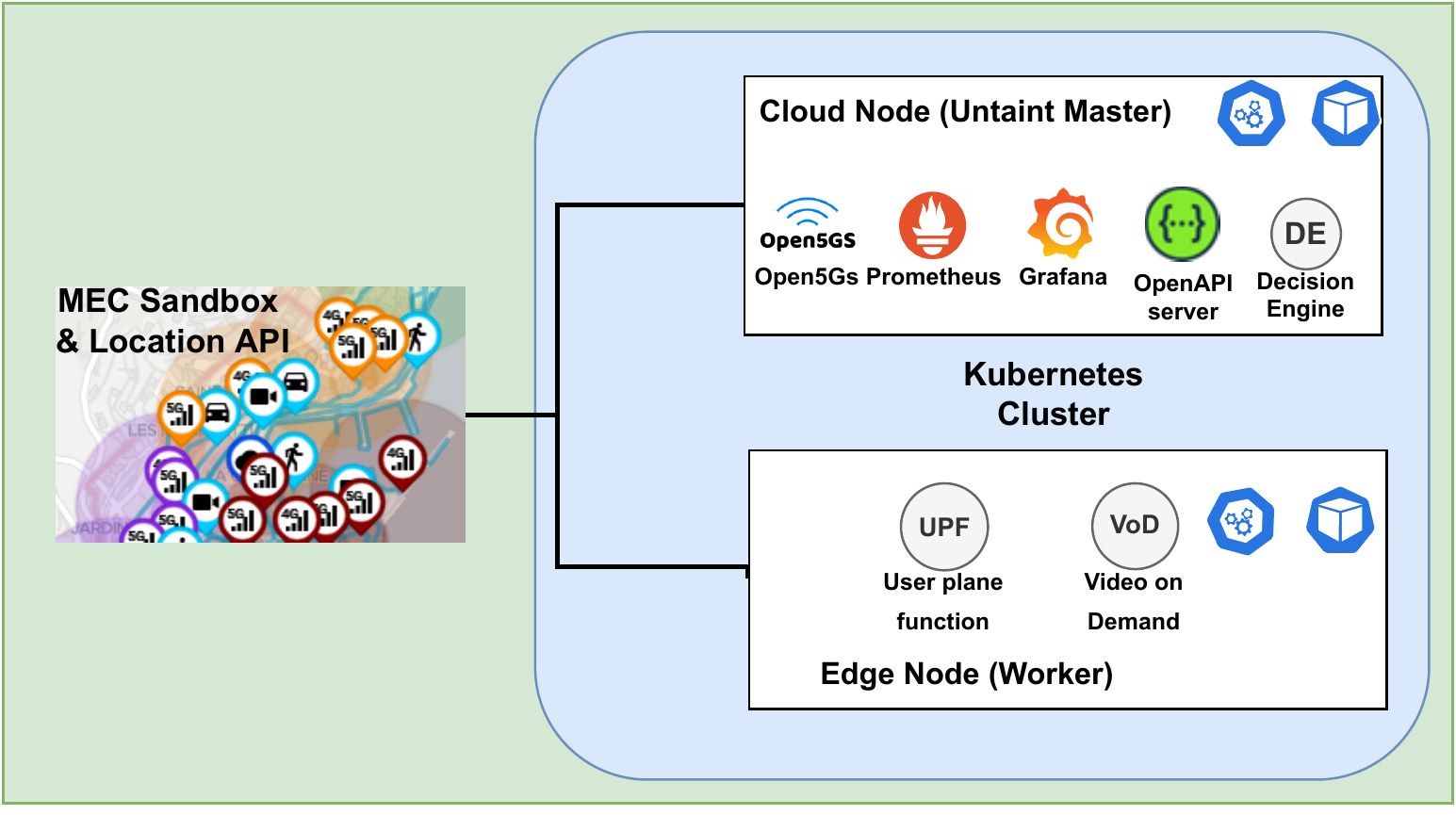}
	\caption{System model}
	\label{fig:shematicoverall}
\end{figure}}

\section{Experiment}
%We use the number of users in zone 3 of the MEC sandbox as a proxy for VoD traffic. In this work, we use threshold-based micro-service scaling. %If the average number of users is greater than $>2$, the decision engine up-scales the VoD application pod from 1 to 2 replicas. If the average number of users goes below $<=2$, the decision engine down-scales the VoD pod from 2 to 1 replica. \sergio{[I propose using variables to make it more formal and adaptable]: 
%If the average number of users is greater than $\Gamma$ (e.g., $\Gamma=2$ in our demo), the decision engine up-scales the VoD application pod accordingly (e.g., from 1 to 2 replicas). If the average number of users goes below $\Gamma$, the decision engine down-scales the VoD pod replicas (e.g., from 2 to 1 replica).}

The experimental setup in MEC Sandbox is configured with the \textit{4g-5g-wifi-macro} scenario, four stationary users, four low-velocity users, and four high-velocity users. The DE calls the MEC Location API to calculate the average number of users of zone 3 and define the threshold to trigger actions through the custom-made Kubernetes OpenAPI.
If the average number of users is greater than $\Gamma$ (e.g., $\Gamma>=3$), the DE upscales the VoD application pod accordingly (e.g., from 1 to 2 replicas). If the average number of users goes below $\Gamma$, the DE downscales the VoD pod replicas (e.g., from 2 to 1 replica).
%If the average number of users is greater than $>3$, the controller upscales the VoD microservice from 1 to 2 replicas. If the average number of users goes below $<=3$, the controller downscales the VoD microservice from 2 to 1 replica. 

\begin{figure}[t]
     \centering
     \begin{subfigure}{.5\textwidth}
        \centering
         \includegraphics[width=.7\textwidth]{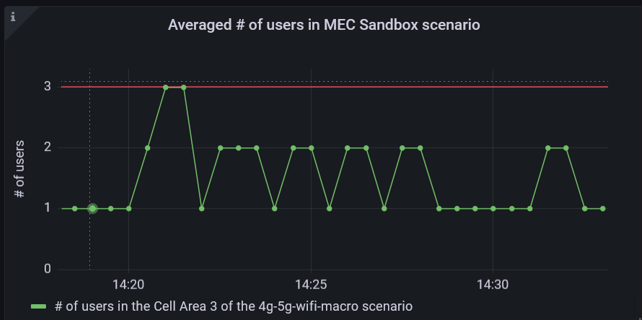}
         \caption{Average number of users}
         \label{fig:user_number}
     \end{subfigure}
     \hfill
     \begin{subfigure}{.5\textwidth}
        \centering
         \includegraphics[width=.7\textwidth]{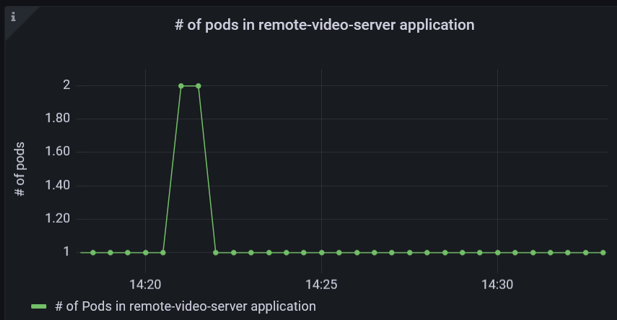}
         \caption{Number of VoD pods}
         \label{fig:pod_number}
     \end{subfigure}

        \caption{Grafana visualization of microservice scaling demo}
        \label{fig:pod_scaling}
\end{figure}

To validate the proper behavior of the DE application, Fig.\ref{fig:user_number} and Fig. \ref{fig:pod_number} illustrate the average number of users and the number of pods, respectively. To visualize the results, Grafana is implemented as another pod alongside Prometheus with the Kube Prometheus Stack helm-chart \cite{kube_stack2022}. The Prometheus server scrapes DE which acts also as a Prometheus exporter and provides Grafana the dataset source to display the metrics. 

%------------------------------------------------------------
%------------------------------------------------------------
\section{Discussion}
The MEC Sandbox does not have any 5G core or RAN deployment. Instead, it provides a coherent API interface that can be used to develop a realistic application. A suitable open-source and cloud-native approach is the Open5GS solution, shown in the system model Fig. \ref{fig:shematicoverall}. Furthermore, the DE is very flexible and can use arbitrary logic for interacting with edge applications like ML-based proactive resource allocation and user trajectory prediction.

The current version of the MEC Sandbox only allows for a total of 12 users, which does not produce enough data for training an ML algorithm. We hope that the future release of the MEC Sandbox resolves this issue and makes it more accessible by open-sourcing the code.

Finally, the proposed framework is fundamentally different from Kubernetes native rule-based scaling (horizontal pod autoscaling). To begin with, we can use service/user-based metrics with the current framework. Also, the framework is not limited to scaling, and the applications running on the edge can directly interact with MEC APIs and RAN, providing added value to both application developers and mobile operators. 

\section{Summary}
In this work, we look at the microservice scaling using MEC Sandbox. The developed decision engine gathers user location information leveraging the ETSI MEC Location API and uses it for microservice scaling in the network edge. The implemented framework can also interact with different MEC APIs, enabling highly configurable application deployment on the mobile edge.  
As a future work, we propose an ML-based DE, which can allocate the edge resources proactively. 
%\michail{In this work, we investigated the scaling of MEC resources from the OpenAPI point of view. The demonstrated DE gathers user location information using the MEC Location API in order to scale a microservice application running in network edge.  }
%------------------------------------------------------------
%------------------------------------------------------------
\section*{Acknowledgment}
The current work is a part of the final submission presented to ETSI / LINUX Foundation Edge Hackathon 2022 by Pedraforca team (Rasoul Nikbakht Silab, Michail Dalgitsis, Sarang Kahvazadeh, Sergio Barrachina) and was supported by the Basque Government under project 
Federated infrastructure of experimentation for Industry 4.0 applications (B-INDUSTRY5G). In addition, this work has been partially funded by the MARSAL project from EU Horizon 2020 research and innovation programme under grant agreement No 101017171, MCIN/AEI/ 10.13039/501100011033 “ERDF A way of making Europe” project under grant PID2021-126431OB-I00, and 5GMediaHuB under agreement no 101016714.
%-----------------------------------------------------------------------------------------------------

\bibliographystyle{IEEEtran}
\bibliography{ref}
\end{document}